\documentclass[twocolumn,prb]{revtex4}

\usepackage{graphicx}
\usepackage{dcolumn}
\usepackage{bm}

\begin{document}

\preprint{APS/123-QED}

\title{Systematic Analysis of Frustration Effects\\in Anisotropic Checkerboard Lattice Hubbard Model}
\author{Takuya Yoshioka$^{1}$}
\author{Akihisa Koga$^{2}$}
\author{Norio Kawakami$^{2}$}
\affiliation{$^{1}$Department of Applied Physics, Osaka University, Suita, Osaka, 565-0871, Japan\\$^{2}$Department of Physics, Kyoto University, Kyoto, 606-8502, Japan}

\date{\today}

\begin{abstract}
We study the ground state properties of the geometrically frustrated Hubbard model on the anisotropic checkerboard lattice with nearest-neighbor hopping $t$ and next nearest-neighbor hopping $t'$. By using the path-integral renormalization group method, we study the phase diagram in the parameter space of the Hubbard interaction $U$ and the frustration-control parameter $t'/t$. Close examinations of the effective hopping, the double occupancy, the momentum distribution and the spin/charge correlation functions allow us to determine the phase diagram at zero temperature, where the plaquette-singlet insulator emerges besides the antiferromagnetic insulator and the paramagnetic metal. Spin-liquid insulating states without any kind of symmetry breaking cannot be found in our frustrated model.

\end{abstract}

\maketitle

\setcounter{footnote}{0}
\section{INTRODUCTION}

Strongly correlated electron systems with frustrated lattice structures have attracted much interest recently. Typical examples are a spinel compound $\rm LiV_2O_4$$^1$ and a pyrochlore compound $\rm Tl_2Ru_2O_7$$^{2,3}$, where the heavy fermion behavior and the Mott transition without magnetic ordering were observed. In these compounds, electron correlations on the frustrated lattice should play a vital role in yielding  a variety of intriguing properties  at low temperatures. Stimulated by the above experimental findings, the Hubbard model on the frustrated pyrochlore lattice and its two-dimensional (2D) analog, called the checkerboard lattice, have been studied intensively.$^{4-30}$

In the previous paper,$^{20}$ we studied the zero-temperature properties of the Hubbard model on the isotropic checkerboard lattice at half filling, and found that the system undergoes a first-order phase transition to the plaquette-singlet insulating (PSI) phase at a finite Hubbard interaction. 
Since the analysis was focused only on the fully frustrated model, 
it is desirable to compare it with less frustrated models, in order to clarify how the frustration affects the nature of the metal-insulator transition. In this connection, we recall that the checkerboard lattice is continuously connected to the ordinary square lattice with electron hopping $t$ by reducing the amplitude of electron hopping $t'$ along diagonal bonds (see Fig. 1). For the square lattice without frustration, it is known that the introduction of infinitesimal Hubbard repulsion induces the metal-insulator transition to the antiferromagnetic insulating (AFI) phase. Therefore, the anisotropic checkerboard lattice, where the amplitude of diagonal hopping $t'$ is modulated to interpolate the above two limiting cases, enables us to clarify the role of frustration in the checkerboard lattice Hubbard model.

\begin{figure}
  \begin{center}
    \begin{tabular}{cc}
      \hspace{-2mm}
      \resizebox{42mm}{!}{\includegraphics{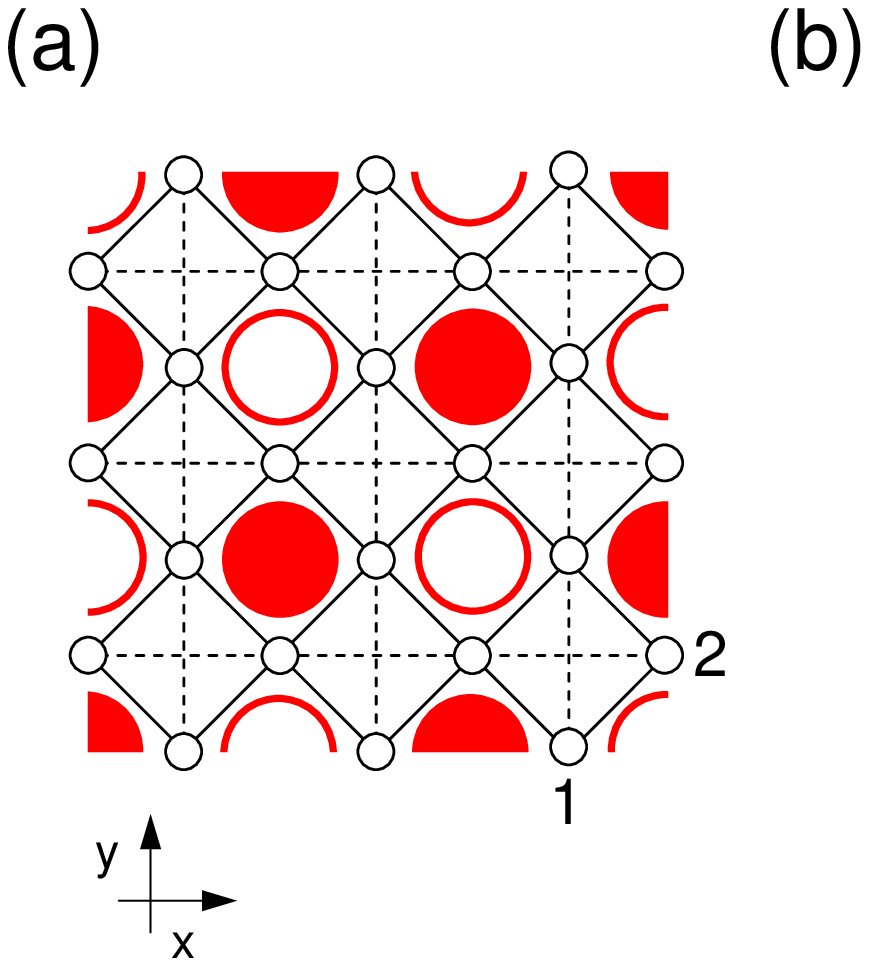}} &
      \hspace{-5mm}
      \resizebox{45mm}{!}{\includegraphics{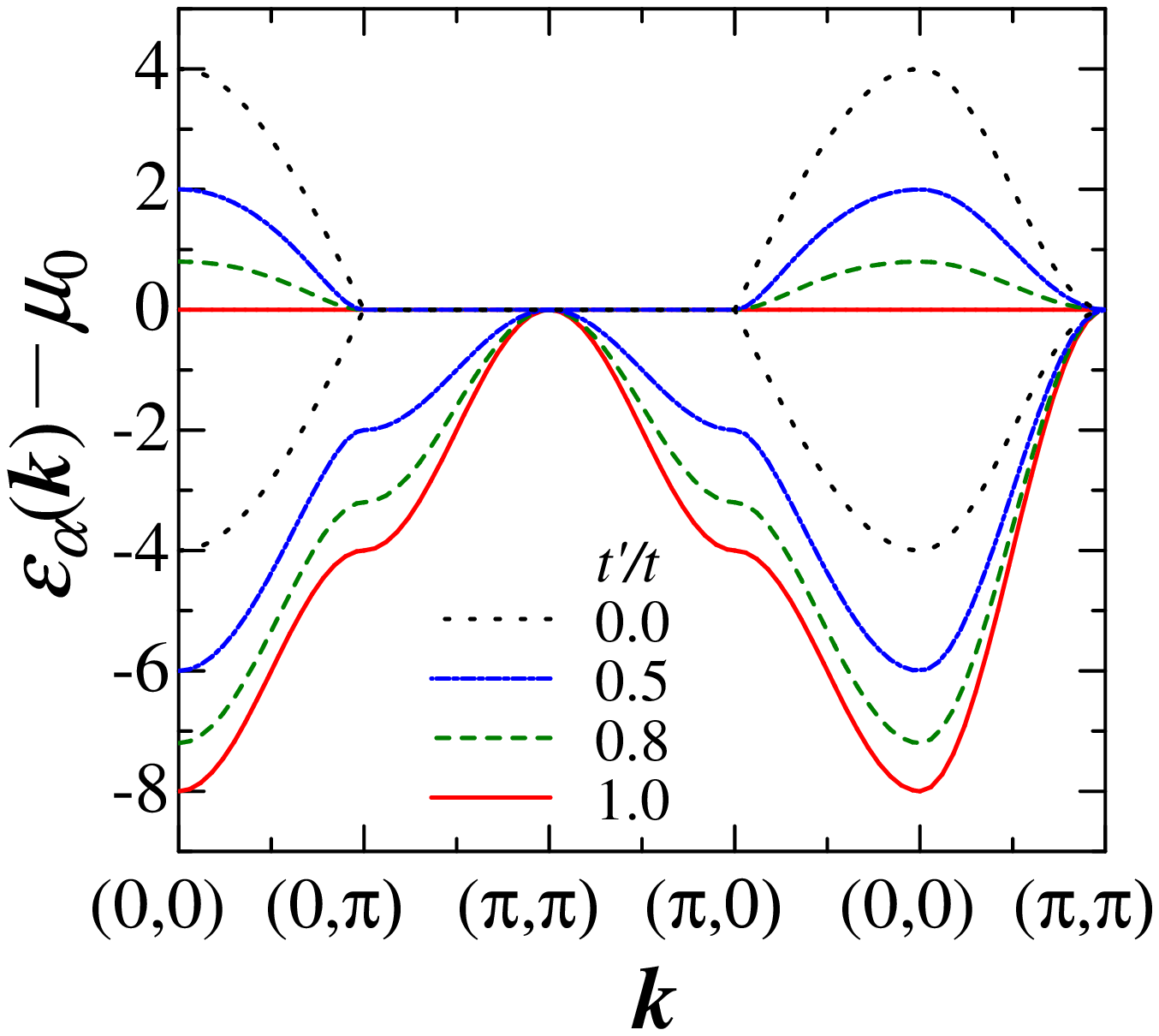}} \\
    \end{tabular}
\caption{(color online). (a) Anisotropic checkerboard lattice. The solid (dotted) lines correspond to the transfer integral $t\ (t')$. Doubly degenerate plaquette valence-bond ordering pattern is shown schematically. Within each unit cell the sublattice sites are denoted by 1,2.
(b) The free electron band structure 
$\varepsilon_\alpha(\mbox{\boldmath$k$})-\mu_0$ 
along symmetry lines in the Brillouin zone (B.z.) for different choices of $t'/t$, where $\mu_0$ is the chemical potential at half filling.
}
\label{fig1}
  \end{center}
\end{figure}

Motivated by this, we here investigate the Hubbard model on the anisotropic checkerboard lattice (Fig. \ref{fig1} (a)) at half filling.  So far, theoretical investigations in this direction have been put forward only in the strong coupling limit, where the half-filled Hubbard model can be mapped to the Heisenberg model with the exchange coupling $J=4t^2/U\ \left(J'=4t'^2/U\right)$. Intensive studies on the spin $1/2$ Heisenberg model on the anisotropic checkerboard lattice$^{31-34}$ concluded that for large (small) $J'/J$, the plaquette valence-bond crystal (AF N\'eel) phase is stabilized, and the corresponding phase transition is of  first order. 
In this paper we aim at establishing the ground-state phase diagram of the anisotropic checkerboard lattice Hubbard model in the wide parameter region from the weak to strong frustration/correlation limit. To this end, we investigate the quantum phase transitions at zero temperature by means of the path-integral renormalization group (PIRG) method developed by Imada group,$^{35-38}$ which is particularly efficient to study electron correlations under strong frustration. We compute the effective hopping, the double occupancy, the momentum distribution and the spin/charge correlation functions. The phase diagram thus determined has the plaquette-singlet insulator besides the antiferromagnetic insulator and the paramagnetic metal. We do not find spin-liquid insulating states without any kind of  symmetry breaking in our phase diagram.

The paper is organized as follows. In \S2, we introduce the model Hamiltonian and briefly explain the PIRG method. We discuss the quantum phase transitions to  obtain the ground-state phase diagram of the anisotropic checkerboard Hubbard model in \S3. A brief summary is given in \S4.

\section{MODEL AND METHOD}
We consider the standard single-band Hubbard model on the anisotropic checkerboard lattice,
\begin{eqnarray}
\hat{\cal H}=-\sum_{i,j,\sigma, (m,m^{'})}t_{ijmm^{'}}
\hat{c}^{\dagger}_{im\sigma}\hat{c}_{jm^{'}\sigma}
+U\sum_{i,m}\hat{n}_{im\uparrow}\hat{n}_{im\downarrow},
\label{H}
\end{eqnarray}
where $\hat{c}_{im\sigma}$ ($\hat{c}^{\dagger}_{im\sigma}$) is an annihilation (creation) operator of an electron in the $i$-th unit cell with spin $\sigma$ and sublattice index $m$ (=1,2), and $\hat{n}_{im\sigma}= \hat{c}^{\dagger}_{im\sigma}\hat{c}_{im\sigma}$. $U$ is the Hubbard repulsion and $t_{ijmm^{'}}(=t,\ t')$ is the transfer integral, which is schematically shown in Fig. \ref{fig1} (a). 
Here, by tuning the ratio of $t'/t$ systematically, we study the ground state properties to clarify how the geometrical frustration affects quantum phase transitions in the system. The kinetic term of the Hamiltonian $\hat{{\cal H}}_k$ is diagonalized at each ${\mbox{\boldmath$k$}}$. Then we obtain $\hat{{\cal H}}_k=\sum_{k,\alpha,\sigma}\varepsilon_{\alpha}({\mbox{\boldmath$k$}})\hat{a}^{\dagger}_{k\alpha\sigma}\hat{a}_{k\alpha\sigma}$ with two eigenvalues,
\begin{eqnarray}
 && \varepsilon_{1,2} ({\mbox{\boldmath$k$}})
 =  t'(\cos k_x+\cos k_y) \\
&&\pm   \sqrt {t'^{2}(\cos k_x-\cos k_y)^2 
 +  16t^2(\cos^2 \frac{k_x}{2})(\cos^2 \frac{k_y}{2})}, \nonumber
\end{eqnarray}
where $+$ and $-$ signs correspond to the bands $\varepsilon_{1}({\mbox{\boldmath$k$}})$ and $\varepsilon_{2}({\mbox{\boldmath$k$}})$. We assume $t,\ t'>0$, hereafter. The dispersion relations thus obtained are shown in Fig. \ref{fig1} (b) along the symmetry lines in the Brillouin zone (B.z.). 

We start by mentioning some characteristics in the two limiting cases of $t'/t=0$ and $t'/t=1$. In  the case of square lattice ($t'/t=0$), it is known that the AFI state is stabilized for any finite $U>0$ at zero temperature, due to the perfect nesting. In our reduced B.z. scheme, the nesting is expressed by the hybridization between upper and lower bands with the same momentum along the boundary of B.z. as shown in Fig. \ref{fig1} (b). On the other hand, for the isotropic checkerboard lattice ($t'/t=1$), the upper band is completely flat over the whole B.z. while the lower band is dispersive at $U=0$. As shown in Refs.$^{14,15}$, the perturbative calculation in $U$ at half filling in the isotropic case suffers from divergence at third and higher orders, because the lower band is completely filled and the Fermi level just touches the flat band. This unusual situation makes the theoretical treatment of the model difficult.


For highly frustrated lattice systems, it is known that powerful quantum Monte Carlo method suffers from the minus sign problem. Also, the exact diagonalization calculation cannot deal with large enough lattice sizes to figure out the role of frustration in our system. To treat the strong correlation and frustration effects, we here make use of the PIRG method$^{35-37}$, where we further employ a quantum number projection (QP) operator to the total spin-singlet state.$^{37}$ The  PIRG+QP method is particularly efficient to study electron correlations under strong frustration. The algorithm is very simple. We start from an unrestricted Hartree-Fock solution$^{39}$ and reach the ground state by taking into account quantum fluctuations in a systematic fashion. We increase the dimension of the truncated Hilbert space in a nonorthogonal Slater basis numerically optimized by the path-integral operation. An energy variance extrapolation$^{35-38,40}$ is very efficient to reach the true ground state of finite size systems. We take the number of Slater basis functions up to 500 and apply an improved iteration scheme proposed in our previous paper$^{20}$. In the present study, we carry out the calculation for the $N=32$ lattice system with periodic boundary conditions. We have already confirmed that the $N=32$ lattice system is large enough to investigate the quantum phase transitions in the thermodynamic limit in the isotropic case$^{20}$.

\section{RESULTS}

To study the quantum phase transitions at zero temperature, we first consider the virtual displacement of $E_g$ with respect to $t'$ and $U$, 
\begin{eqnarray}
\delta E_g(t',U)
&=&\left(\frac{\partial E_g}{\partial t'}\right)_U\delta t'+
\left(\frac{\partial E_g}{\partial U}\right)_{t'}\delta U,
\end{eqnarray}
and evaluate each coefficient. The former coefficient describes the averaged hopping amplitude along diagonal bonds, while the latter the double occupancy of electrons at each site. Both quantities provide important information about the quantum phase transition. 

To discuss the spin properties, we also calculate the site-dependent spin correlation function $C_{t(t')}$ defined by
\begin{eqnarray}
C_t
&=&\frac{1}{N}\frac{1}{N_t}
\sum_{i=1}^{N}\sum_{\tau_t=1}^{N_t}
\langle 
\hat{\mbox{\boldmath$S$}}_{i}\cdot 
\hat{\mbox{\boldmath$S$}}_{i+\tau_t}\rangle,
\end{eqnarray} 
where $\tau_t$ labels neighboring sites connected by the transfer integral $t$ and $N_t$ is the number of them ($N_t=4$, $N_{t'}=2$). $C_{t'}$ is obtained by replacing $\tau_t$ with $\tau_{t'}$ in the formula (4).
For our anisotropic checkerboard lattice, the plaquette spin singlet state is one of the most probable candidates for the ground state, so that we examine the following plaquette correlation function $P_{A(B)}$ defined by
\begin{equation}
\begin{array}{rcl}
\displaystyle
P_{A}
&=&
\displaystyle
\langle \hat{Q}_{A}^2\rangle, \label{P}\vspace{2mm}
\\
\displaystyle
\hat{Q}_{A}&=&
\displaystyle
\frac{2}{N}
\sum_{\stackrel{i=1}{{\rm pattern}\ A}}^{N/2}(-1)^i\hat{p}_i,
\end{array}
\end{equation}
where $\hat{p}_i[=
(\hat{\mbox{\boldmath$S$}}_{\alpha_i}+
\hat{\mbox{\boldmath$S$}}_{\gamma_i})\cdot (
\hat{\mbox{\boldmath$S$}}_{\beta_i}+
\hat{\mbox{\boldmath$S$}}_{\delta_i})]$ is the $i$th plaquette-singlet operator. The corresponding configuration pattern of plaquettes and their signs $(-1)^i$ are schematically shown in Fig. \ref{fig1} (a), where filled (open) circles represent positive (negative) signs for $P_{A}$. The pattern for the correlation $P_{B}$ is given by totally shifting the circles to the square with crossing in Fig. \ref{fig1} (a).  Note here that in  the autocorrelation terms $\sum_i\langle \hat{p}_i^2\rangle$ in Eq. (5) equally contribute to $P_A$ and $P_B$ in the AF ordered phase, so that the plaquette correlation functions $P_A$ and $P_B$ could have finite values even in this phase. Therefore, we also examine the value of $P_{A}-P_{B}$ in order to distinguish the plaquette singlet phase and the AF phase.

\begin{figure}[htb]
  \begin{center}
\includegraphics[width=0.45\textwidth]{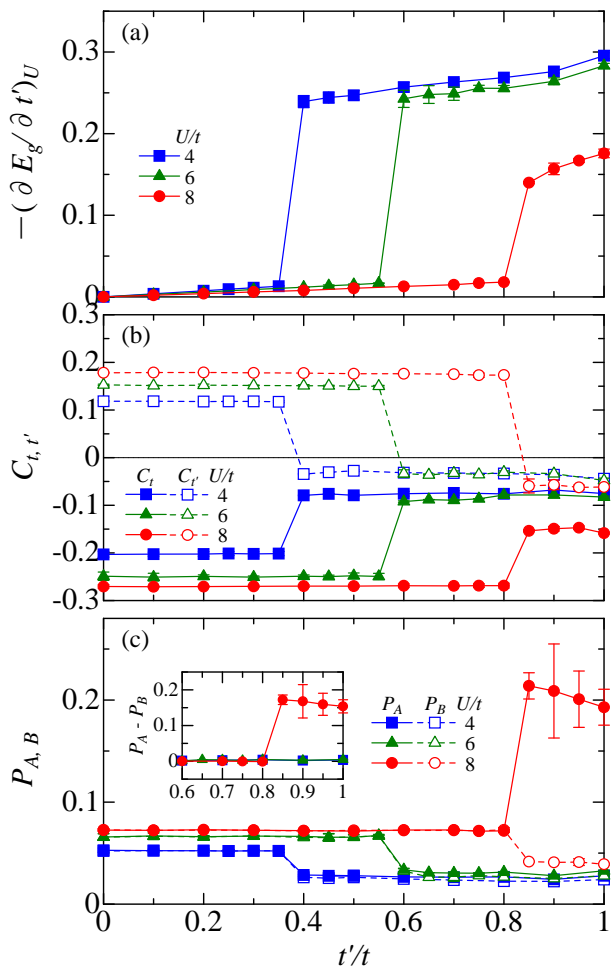}
\caption{
(color online). (a) The averaged hopping, $-\left(\partial E_g/\partial t'\right)_{U}$, (b) the spin correlation $C_{t(t')}$, and (c) the plaquette correlation $P_{A(B)}$ as a function of $t'/t$ on the $N=32$ lattice system at half filling for different choices of $U/t$. The inset of (c) shows $P_A-P_B$.
}
\label{fig2}
  \end{center}
\end{figure}

\subsection{Phase transitions under frustration control}

In the following, we present the computed results in two ways; quantum phase transitions are discussed under control of (i) frustration and (ii) electron correlations. In this subsection,  we first address the frustration control, and show the above physical quantities as a function of $t'/t$ for different choices of $U/t$. The results are summarized in Fig. \ref{fig2}. Starting from the square lattice with $t'=0$, we examine the instability of the AFI state which is stabilized at $t'=0$ for any finite $U/t>0$. In the AFI phase, two electron spins sitting on the nearest neighbor (next nearest neighbor) sites have AF (ferromagnetic) correlations, as indeed confirmed in Fig. \ref{fig2}(b).  Correspondingly, the absolute value of the averaged hopping on the $t'$ bond, $\left(\partial E_g/\partial t'\right)_U$, is strongly suppressed in the AFI phase (Fig. \ref{fig2} (a)). These results indicate the presence of AF order in the small $t'/t$ region at least for the Hubbard repulsion, $U \geq 4$. We also note that the double occupancy $\left(\partial E_g/\partial t'\right)_U$ is almost unchanged in the AFI phase even if $t'/t$ is altered (though not shown in the figure). When the effect of frustration is further enhanced via increase of $t'/t$, the AFI state becomes unstable, triggering a phase transition to another insulating phase. The transition point depends on the strength of the Hubbard $U$. We indeed observe the abrupt jump in  $\left(\partial E_g/\partial t'\right)_U$ at $t'/t\simeq 0.4$, $0.6$, and $0.8$ for $U/t=4,\ 6$, and $8$, respectively (Fig. \ref{fig2}(a)). Similar discontinuities in the two types of correlation functions are found at the same transition points in Figs. \ref{fig2} (b) and (c). Therefore we conclude that the phase transition is of first order. 

In order to see the nature of the insulating phase realized at larger $t'/t$ in  detail, we look at the spin correlation $C_{t(t')}$ and the plaquette correlation $P_{A(B)}$. In the case of $U/t=4$ and $6$, we find that the first-order transition  occurs between AFI and PM, as clearly seen in $C_{t(t')}$ shown in Fig. \ref{fig2} (b). 
In these cases, once the transition occurs $P_{A}$ and $P_{B}$ are  both reduced and the value of $P_{A}-P_{B}$ is almost zero, implying that the plaquette-singlet state is not formed. We also note that in the  PM phase  $-\left(\partial E_g/\partial t'\right)_U$ has the value about 0.3, which is a little bit smaller than $1/3$ expected for the noninteracting case (Fig. \ref{fig2} (a)). 
 
In contrast to the above two cases, quite different behavior emerges in the correlation function in the case of $U/t=8$: not only the magnitude of $P_{A(B)}$ itself but also $P_{A}-P_{B}$ are abruptly increased at $t=t'_c$, as seen in Fig. \ref{fig2} (c) and its inset, implying that the transition from AFI to PSI indeed occurs. We have confirmed that the same type of transition occurs at $t'/t\simeq 0.9$ for $U/t=10$. In the strong coupling limit with large $U$, we can check how precise our estimate of the transition point is in terms of the effective model. In this limit, our system is mapped to the anisotropic checkerboard Heisenberg model with two different exchange couplings $J$ and $J'$. The value of $t'_c/t$ for the transition point then yields  $J'_c/J\simeq0.8$, which agrees very well with $J'_c/J=0.79-0.81$ for the Heisenberg model estimated by a strong-coupling expansion$^{33}$. The good agreement confirms the validity of our analysis, and in turn supports the existence of the first order phase transition between AF N\'eel and plaquette valence-bond crystal phase proposed for the Heisenberg model$^{32-34}$. We note here that the above results are totally consistent with our previous study of the isotropic model ($t'/t=1$), where much more systematic analyses, performed with finite-size scaling, give $U/t=6.75\pm 0.25$$^{20}$ for the PM-PSI transition point$^{20}$.

\subsection{Phase transitions under control of electron correlations}

\begin{figure}[htb]
  \begin{center}
\includegraphics[width=0.45\textwidth]{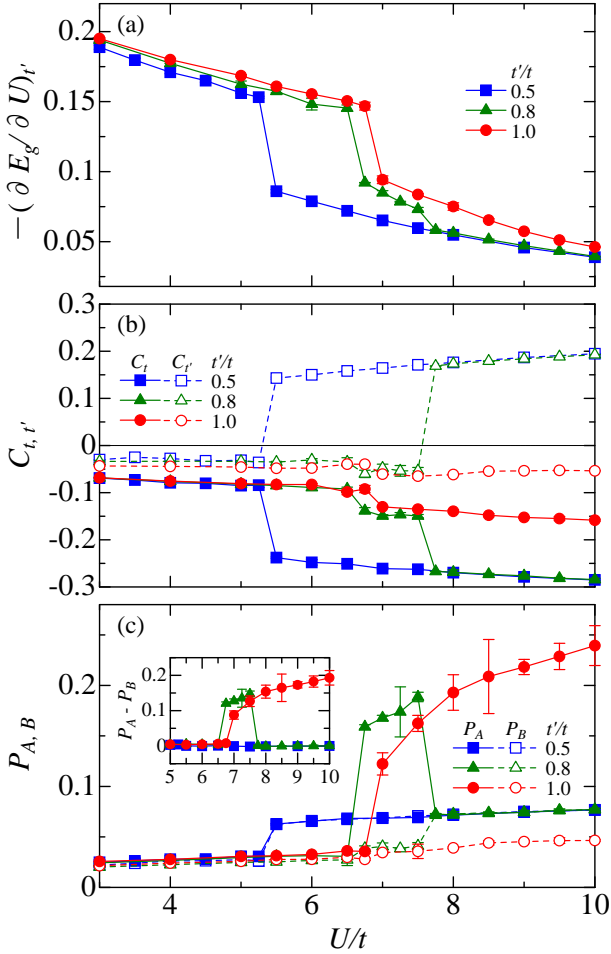}
\caption{
(color online). (a) The double occupancy $\left(\partial E_g/\partial U\right)_{t'}$, (b) the spin correlation $C_{t(t')}$, and (c) the plaquette correlation $P_{A(B)}$ as a function of $U/t$ on the $N=32$ lattice system at half filling for different choices of $t'/t$. The inset of (c) shows $P_A-P_B$.
}
\label{fig3}
\end{center}
\end{figure}

We next show the physical quantities in Eqs. (3)-(5) as a function of $U/t$ for different choices of $t'/t$ to discuss the quantum phase transitions under control of electron correlations. We show the $U/t$-dependence of the double occupancy $\left(\partial E_g/\partial U\right)_{t'}$ in Fig. \ref{fig3} (a). The introduction of the Hubbard interaction monotonically decreases the double occupancy, implying that the paramagnetic metallic state is realized in the small $U$ region. Further increase in the interaction yields the discontinuity of $\left(\partial E_g/\partial U\right)_{t'}$, in accordance with the first order Mott transition. We determine the transition point $U_c$  by estimating the level crossing point of energies for the competing metallic and insulating states. 

\begin{figure*}
  \begin{center}
    \begin{tabular}{cp{3mm}cccp{3mm}p{60mm}}
      \vspace{5mm}
      &&$U/t=5$&$U/t=7$&$U/t=9$&&\\
      \vspace{5mm}
      \raisebox{13mm}{$n_{\alpha}({\mbox{\boldmath$k$}})$}
      &&\includegraphics[width=30mm]{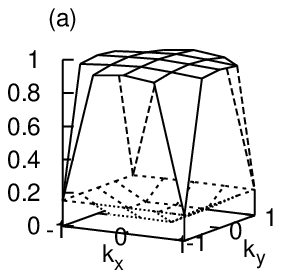} &
      \includegraphics[width=30mm]{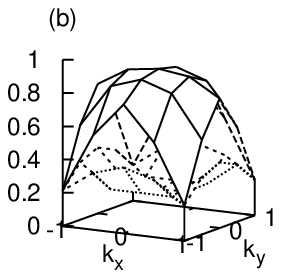} &
      \includegraphics[width=30mm]{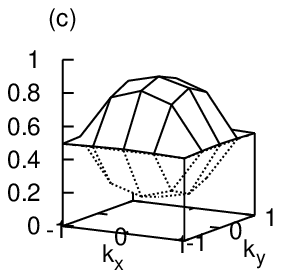} &&\\
      \vspace{5mm}
      \raisebox{13mm}{$N_{\beta}({\mbox{\boldmath$q$}})$}
      &&\includegraphics[width=30mm]{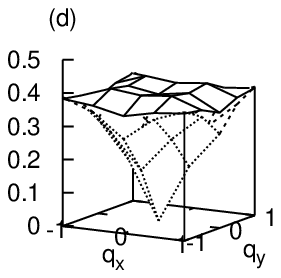} &
      \includegraphics[width=30mm]{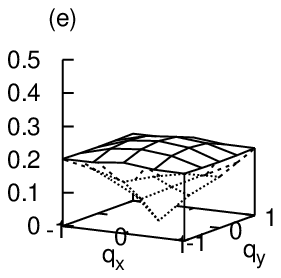} &
      \includegraphics[width=30mm]{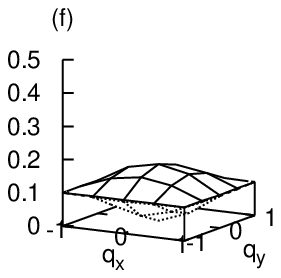} &&\\
      \raisebox{13mm}{$S_{\beta}({\mbox{\boldmath$q$}})$}
      &&\includegraphics[width=30mm]{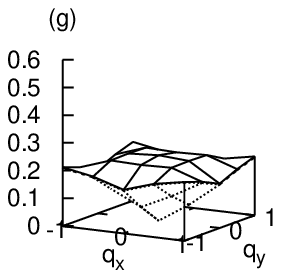} &
      \includegraphics[width=30mm]{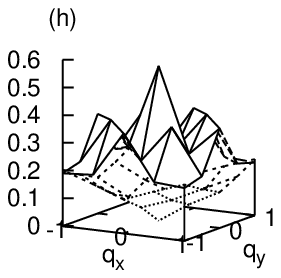} &
      \includegraphics[width=30mm]{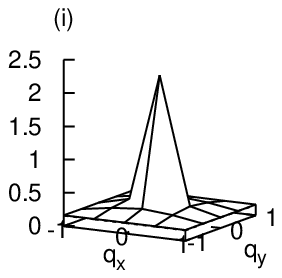} &&\vspace{-70mm}\caption{(a)-(c) The momentum distribution function $n_\alpha ({\mbox{\boldmath$k$}})$, where $\alpha=1\ (2)$ corresponds to the dashed (solid) lines and (d)-(f) [(g)-(i)] the momentum-dependent charge [spin] correlation function $N_{\beta}({\mbox{\boldmath$q$}})$ [$S_{\beta}({\mbox{\boldmath$q$}})$], where $\beta=$max (min) corresponds to the solid (dashed) lines at $t'/t=0.8$ for different choices of $U/t$. Note that $4S_{\beta}({\mbox{\boldmath$q$}})=N_{\beta}({\mbox{\boldmath$q$}})$ at $U/t=0$.}\\
    \end{tabular}
  \end{center} 
    \label{fig4}
\end{figure*}

In the case of $t'/t=0.5$, $C_t$ and $C_{t'}$ both show discontinuities at $U_c/t\simeq 5.4$ and suddenly increase with opposite signs, suggesting the transition to the AFI phase. For larger $U\ (>U_c)$, the absolute value of $C_t$ is smaller than that for $C_t=-0.335$ known for the square lattice Heisenberg model ($J'/J=0$)$^{41}$. The difference comes from the presence of charge fluctuations which reduce the local magnetic moment. For $t'/t=0.8$, the phase transitions take place twice at $U_{c1}/t\simeq 6.6$ and $U_{c2}/t\simeq 7.6$, and the PSI phase is realized for $U_{c1}<U<U_{c2}$ between the PM and AFI phases (see Fig. \ref{fig3} (c) and its inset). Therefore we can see three types of phase transitions under control of correlations: a single Mott transition of PM-AFI ($t'/t=0.5$) and PM-PSI ($t'/t=1$), and double quantum phase transitions of PM-PSI and PSI-AFI ($t'/t=0.8$).

To investigate the nature of the quantum phase transitions in detail, we further calculate the momentum distribution $n_\alpha ({\mbox{\boldmath$k$}})$ and the momentum-dependent correlation functions in the charge [spin] sector $N_{mm^{'}}({\mbox{\boldmath$q$}})$ $[S_{mm^{'}}({\mbox{\boldmath$q$}})]$ at $t'/t=0.8$, which are given by
\begin{equation}
n_\alpha ({\mbox{\boldmath$k$}})
=\left\{
\begin{array}{ll}
\displaystyle
\frac{1}{2N}\sum_{\sigma}\langle \hat{a}^{\dagger}_{k\alpha\sigma}
\hat{a}_{k\alpha\sigma}\rangle\ &{\rm for}\ {\mbox{\boldmath$k$}}\neq(\pi,\pi), \\
\displaystyle
\frac{1}{4N}\sum_{\sigma,\beta}\langle \hat{a}^{\dagger}_{k\beta\sigma}
\hat{a}_{k\beta\sigma}\rangle\ &{\rm for}\ {\mbox{\boldmath$k$}}=(\pi,\pi),
\end{array}
\right.
\end{equation}
\begin{eqnarray}
N_{mm^{'}}
({\mbox{\boldmath$q$}})
&=&\frac{2}{N}
\sum_{i,j=1}^{N/2}
\left(
\langle \hat{n}_{im}\hat{n}_{jm^{'}}\rangle-\langle \hat{n}_{im}\rangle\langle \hat{n}_{jm^{'}}\rangle
\right)\nonumber\\ 
&\times& e^{i{\mbox{\boldmath$q$}}\cdot 
(\mbox{\boldmath$R$}_{im}-\mbox{\boldmath$R$}_{jm^{'}})},\\
S_{mm^{'}}
({\mbox{\boldmath$q$}})
&=&\frac{2}{3N}
\sum_{i,j=1}^{N/2}
\langle 
\hat{\mbox{\boldmath$S$}}_{im}\cdot 
\hat{\mbox{\boldmath$S$}}_{jm'}
\rangle\nonumber\\ 
&\times& 
e^{i{\mbox{\boldmath$q$}}\cdot 
(\mbox{\boldmath$R$}_{im}-\mbox{\boldmath$R$}_{jm^{'}})},
\end{eqnarray}
where $\hat{n}_{im}=\hat{n}_{im\uparrow}+\hat{n}_{im\downarrow}$ and $\mbox{\boldmath$R$}_{im}$ represents the position of the  $i$-th unit cell in the $m$-th sublattice. Diagonalizing the $2\times2$ matrix, we obtain $S_{\beta}({\mbox{\boldmath$q$}})$ and $N_{\beta}({\mbox{\boldmath$q$}})$ $(\beta={\rm max,\ min})$.

\begin{figure}[htb]
  \begin{center}
\includegraphics[width=0.45\textwidth]{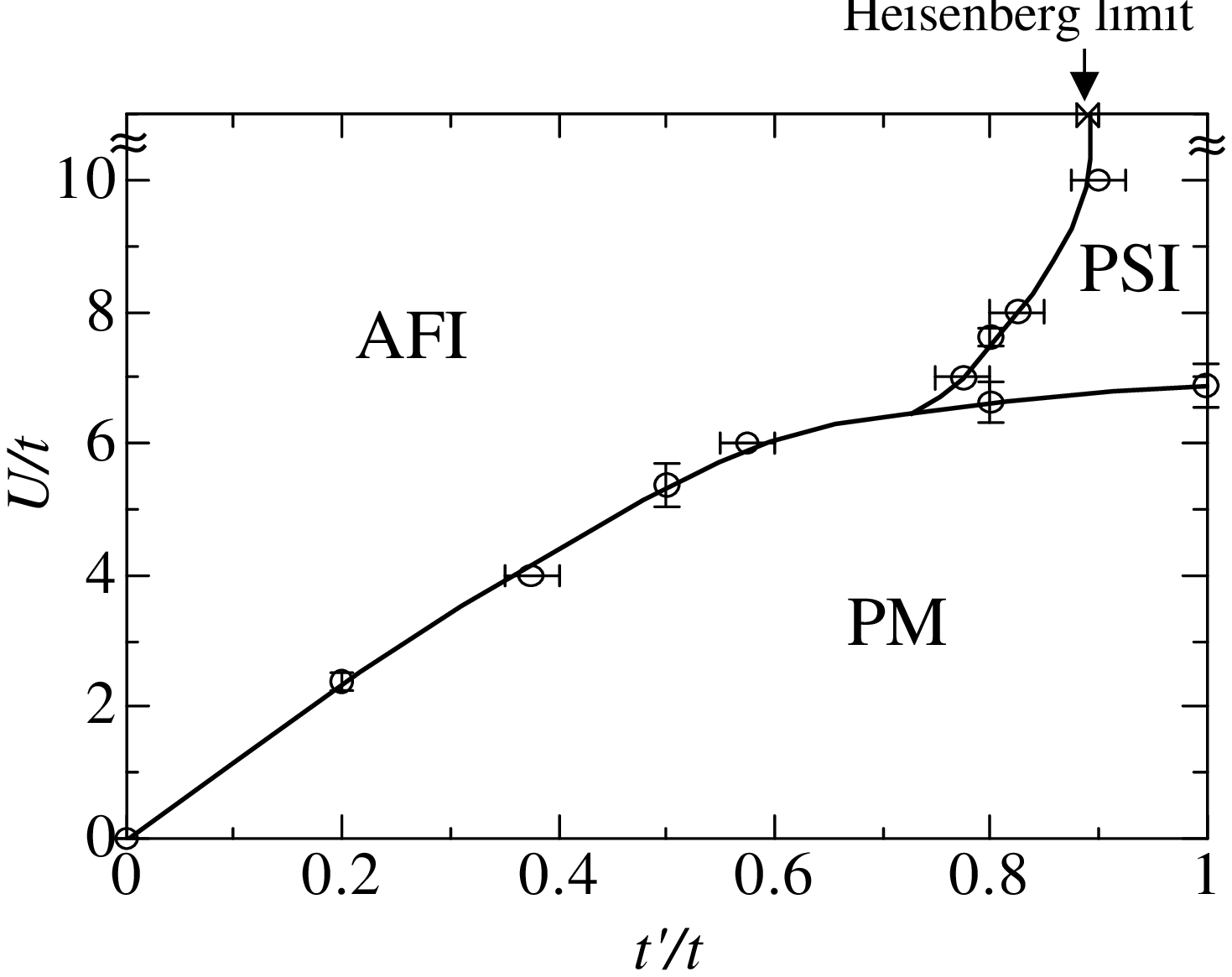}
\caption{
The ground-state phase diagram of the anisotropic checkerboard lattice Hubbard model. AFI, PSI, and PM represent the  antiferromagnetic insulator, the plaquette singlet insulator and the paramagnetic metal, respectively. Calculated results for the $N=32$ system are shown by open circles with error bars. The phase transition point is determined by the crossing point of the energies for two competing phases, and its error bars originate from the extrapolation procedure of the energies. Cross symbol indicates the phase boundary of the $U\rightarrow \infty$ Heisenberg limit$^{33}$ (see text). The quantum phase transitions are of first-order.  
}
\label{fig5}
  \end{center}
\end{figure}

We show the computed results in Fig. 4. In the PM phase, the quasi-Fermi surface exists and therefore $n_{2}({\mbox{\boldmath$k$}})$ has a discontinuity at ${\mbox{\boldmath$k$}}=(\pi,\pi)$ for $U\leq U_{c1}$ (Fig. 4 (a)). As $U$ increases, the PM state becomes unstable and then the discontinuity of $n_{2}({\mbox{\boldmath$k$}})$ disappears at $U\geq U_{c1}$  in the insulating phase (Fig. 4 (b)-(c)). Moreover we can confirm that the Hubbard interaction reduces the charge fluctuations so that $N_{\beta}({\mbox{\boldmath$q$}})$ is uniformly reduced over the whole B.z. (Fig. 4 (d)-(f)). By contrast, the spin correlations $S_{\rm max}({\mbox{\boldmath$q$}})$ are totally enhanced. Especially a peak structure is developed for $U\geq U_{c1}$ at ${\mbox{\boldmath$q$}}=(0,0)$ (Fig. 4 (h)-(i)), while there is no such peak structure for $U\leq U_{c1}$ (Fig. 4 (g)). Also, divergent increase in $S_{\rm max}(0,0)$ is observed beyond a certain interaction $U_{c2}$. Therefore the phase transition between PSI and AFI occurs at $U=U_{c2}$. As a result of the transition, the momentum distributions $n_1({\mbox{\boldmath$k$}})$ and $n_2({\mbox{\boldmath$k$}})$ have the reflection symmetry with respect to the $n_{\alpha}({\mbox{\boldmath$k$}})=0.5$ plane  (Fig. 4 (c)). These properties in the AFI phase should be adiabatically connected to those in the SDW phase for the square lattice Hubbard model ($t'/t=0$). 

Summarizing all the above results, we end up with the phase diagram of  the Hubbard model on the anisotropic checkerboard lattice, as shown in Fig. \ref{fig5}. There are three distinct phases of PM,  AFI and PSI. The quantum phase transitions among them are of first order.  For small $U$, the PM phase appears, while for large $U$ and small $t'$, the AFI phase is realized in accordance with the known results. The PSI phase is stabilized in the strong frustration region with $t'/t \sim 1$ and large $U$. Note that around $t'/t \sim 0.8$ and $U/t \sim 6$, the three phases strongly compete with each other. In fact, the double phase transitions, which occur around $t'/t=0.8$ as $U$ increases, reflect this kind of strong competition. We would like to stress again that in the two limiting cases with strong frustration, the phase boundary obtained here reproduce the known results fairly well: the Heisenberg limit$^{33}$ with large $U$  and the isotropic checkerboard limit$^{20}$  with $t'/t=1$, both of which were studied in detail previously. We therefore believe that the phase diagram obtained in this paper is reliable although we have restricted our analysis to the $N=32$ lattice system.

\section{SAMMARY}

We have studied the ground-state properties of the anisotropic checkerboard lattice Hubbard model by means of the PIRG method. By controlling the geometrical frustration via a systematic change in the transfer integral $t'$ along diagonal bonds, we have dealt with the wide parameter region from the square lattice ($t'=0$) to the fully-frustrated isotropic checkerboard lattice ($t'/t=1$). The ground state phase diagram thus obtained consists of three distinct quantum phases depending on the Hubbard interaction $U$ and the strength of frustration $t'/t$. 

In particular, in the region with strong frustration ($t'/t \sim 1$), we have the plaquette singlet insulator with broken translational symmetry, which is in contrast to the results for analogous two-dimensional frustrated electron systems such as the anisotropic triangular lattice model where a quantum spin liquid phase without any symmetry breaking was proposed for the ground state$^{42-49}$. Therefore, it remains an important problem to figure out what is really relevant for realizing the insulating phase without symmetry breaking. It is also interesting to investigate the nature of the finite-temperature Mott transition of the anisotropic checkerboard lattice. In particular, it is worth exploring whether the reentrant behavior in the temperature-driven Mott transition found for the anisotropic triangular lattice model$^{50}$ could emerge in the checkerboard lattice model. These issues are now under consideration.

\begin{acknowledgments}
This work was partly supported by the Grant-in-Aid for Scientific Research 
[19014013, 20029013 (N.K.) and 20740194 (A.K.)] and the Global COE Program "The Next Generation of Physics, Spun from Universality and Emergence" from the Ministry of Education, Culture, Sports, Science and Technology (MEXT) of Japan. A part of computations was done at the Supercomputer Center at the Institute for Solid State Physics, University of Tokyo and Yukawa Institute Computer Facility. T. Y. is supported by the Japan Society for the Promotion of Science. 
\end{acknowledgments}

\end{document}